\documentclass[twocolumn,twoside,slac_two]{revtex4}
\usepackage{graphicx}
\usepackage{fancyhdr}
\newcommand{\be}{\begin{equation}}
\newcommand{\ee}{\end{equation}}
\newcommand{\bea}{\begin{eqnarray}}
\newcommand{\eea}{\end{eqnarray}}
\begin{document}

\title{Cosmic ray transport in MHD turbulence}

\author{Huirong Yan and A. Lazarian}
\affiliation{Department of Astronomy, University of Wisconsin-Madison}

\begin{abstract}
Recent advances in understanding of magnetohydrodynamic (MHD) turbulence
call for revisions in the picture of cosmic ray transport.
In this paper we use recently obtained scaling laws for MHD modes
to obtain the scattering frequency for cosmic rays. 
We account for the turbulence cutoff arising from both collisional and collisionless
damping. We obtain the scattering rate and show that fast modes
provide the dominant contribution to cosmic ray scattering for the
typical interstellar conditions in spite of the fact that fast modes
are subjected to damping. We determine how the efficiency of the scattering depends on the characteristics of ionized media, e.g. plasma $\beta$. We show that streaming instability is suppressed by the ambient MHD turbulence. 
\end{abstract}

\maketitle

\thispagestyle{fancy}

\section{Introduction}

The propagation of cosmic rays (CRs) is affected by their interaction
with magnetic field. This field is turbulent and therefore, the resonant
interaction of cosmic rays with MHD turbulence has been discussed
by many authors as the principal mechanism to scatter and isotropize
cosmic rays. Although cosmic ray diffusion can
happen while cosmic rays follow wandering magnetic fields (\cite{Jokipii}), the acceleration of cosmic rays requires efficient scattering.
For instance, scattering of cosmic rays back into the shock is a
vital component of the first order Fermi acceleration.

While most investigations are restricted to Alfv\'{e}n modes propagating
along an external magnetic field (the so-called slab model of Alfv\'{e}nic
turbulence), obliquely propagating MHD modes have been included in
\cite{Fisk} and later studies \cite{PP}. A more complex models were obtained by combining the 
results of the Reduced MHD with parallel slab-like
modes have been also considered \cite{Bieber}.
Here we attempt to use models that are motivated by
the recent studies of MHD turbulence 
(\cite{GS95}, see \cite{CLY} for a 
review and references therein).  

The efficiency of scattering depends on turbulence anisotropy \cite{Lerche}. Therefore the calculations
of CR scattering must be done using a realistic MHD turbulence model.
An important attempt in this direction was carried out in \cite{Chan}. However, only incompressible motions were considered. On the 
contrary, ISM is highly compressible. Compressible MHD turbulence has been studied recently (see review by \cite{CL03} and references therein).
\cite{Sch98} addressed the scattering by fast modes.
But they did not consider the damping, which is essential for fast
modes. In this paper we include various damping processes
which can affect the fast modes. 

\section{MHD turbulence and its damping}

\subsection{Model of MHD turbulence}

Analogous to MHD perturbations that can be decomposed into Alfv\'{e}nic, slow and fast
waves with well-defined dispersion relations, MHD perturbations that characterize turbulence 
can be separated into distinct modes.
The separation into Alfv\'en and pseudo-Alfv\'en modes, which
are the incompressible limit of slow modes, is an essential element  
of the GS95 \cite{GS95} model.
Even in a compressible medium, MHD turbulence is not an inseparable mess in spite of the fact that MHD 
turbulence is a highly non-linear phenomenon \cite{LG01,CL02}. 
The actual decomposition of MHD turbulence into Alfv\'en, slow and fast 
modes was addressed in \cite{CL02,CL03}, who
used a statistical procedure of
decomposition in the Fourier space, where the basis of the Alfv\'en, slow
and fast perturbations was defined. 

Unlike hydrodynamic turbulence, Alfv\'{e}nic one is anisotropic, with 
eddies elongated
along the magnetic field. On the
intuitive level it can be explained as the result of the following fact:
it is easier to mix the magnetic field lines perpendicular
to the direction of the magnetic field rather than to bend them. 
However, one cannot do mixing in the perpendicular direction to very
small scales without affecting the parallel scales. This is probably the
major difference between the adopted model of Alfv\'enic perturbations and
the Reduced MHD \cite{Bieber94}. In the GS95 model as well as in
its generalizations for compressible medium mixing motions induce
the reductions of the scales of the parallel perturbations. 

The corresponding
scaling can be easily obtained. For instance, calculations in \cite{CLV02}
prove that motions perpendicular to magnetic field lines are essentially
hydrodynamic. As the result, energy transfer rate due to those motions
is constant $\dot{E_{k}}\sim v_{k}^{2}/\tau_{k}$, where $\tau_{k}$
is the energy eddy turnover time $\sim(v_{k}k_{\perp})^{-1}$, where
$k_{\perp}$ is the perpendicular component of the wave vector $\mathbf{k}$.
The mixing motions couple to the wave-like motions parallel to magnetic
field giving a critical balance condition, i.e., $k_{\bot}v_{k}\sim k_{\parallel}V_{A}$,
where $k_{\parallel}$ is the parallel component of the wave vector
$\mathbf{k}$, $V_{A}$ is the Alfv\'en speed%
\footnote{note that the linear dispersion relation is used for Alfv\'en modes.
}. From these arguments,
the scale dependent anisotropy $k_{\parallel}\propto k_{\perp}^{2/3}$and
a Kolmogorov-like spectrum for the perpendicular motions 
$v_{k}\propto k^{-1/3}$ can be obtained \cite{LV99}.

It was conjectured in \cite{LG01} that GS95 scaling
should be approximately true for Alfv\'{e}n and slow modes in moderately
compressible plasma. For magnetically dominated, the so-called low
$\beta$ plasma, \cite{CL02} showed that the coupling of Alfv\'{e}nic and
compressible modes is weak and that the Alfv\'{e}nic and slow modes
follow the GS95 spectrum. This is consistent with the analysis of
HI velocity statistics \cite{LP, SL} as well as with the electron density statistics \cite{Armstrong}. Calculations in \cite{CL03} demonstrated that fast modes are marginally
affected by Alfv\'{e}n modes and follow acoustic cascade in both
high and low $\beta$ medium. In what follows, we consider both Alfv\'{e}n
modes and compressible modes and use the description of those modes
obtained in \cite{CL02, CL03} to study CR scattering by MHD turbulence.

\subsection{Damping of Turbulence}

In many earlier papers Alfv\'{e}nic turbulence
was considered by many authors as the default model of interstellar
magnetic turbulence. This was partially motivated by the fact that
unlike compressible modes, the Alfv\'{e}n ones are essentially free
of damping in fully ionized medium\footnote{This picture contradicts to an erroneous assumption
of strong coupling of compressible and incompressible MHD modes
that still percolates the literature on turbulent star formation (see
discussion in \cite{LC}).
However, little cross talk between the different astrophysical
communities allowed these two different pictures to coexist peacefully.}. 
However, it was shown that compressible fast modes are
particularly important for cosmic ray scattering \cite{YL02, YL04}. For them damping is essential.

At small scales turbulence spectrum is altered by damping. Various
processes can damp the MHD motions (see \cite{YL04} for details).
In partially ionized plasma, the ion-neutral collisions are the dominant
damping process. In fully ionized plasma, there are basically two
kinds of damping: collisional or collisionless damping. Their relative
importance depends on the mean free path 
in the medium. If the wavelength is larger than
the mean free path, viscous damping dominates. If, on the other hand,
the wavelength is smaller than mean free path, then the plasma is
in the collisionless regime and collisionless damping is dominant.

To obtain the truncation scale, the damping time $\Gamma_{d}^{-1}$
should be compared to the cascading time $\tau_{k}$. As we mentioned
earlier, the Alfv\'{e}nic turbulence cascades over one eddy turn
over time $(k_{\perp}v_{k})^{-1}\sim(k_{\parallel}V_{A})^{-1}$. The
cascade of fast modes takes a bit longer: 

$$
\tau_{k}=\omega/k^{2}v_{k}^{2}=(k/L)^{-1/2}\times V_{ph}/V^{2},
\label{fdecay}
$$
where $V$ is the turbulence velocity at the injection scale, $V_{ph}$ is is the phase speed of fast modes and equal to Alfv\'en and sound velocity for high and
low $\beta$ plasma, respectively \cite{CL02}. If the damping is faster than
the turbulence cascade, the turbulence gets
 truncated. Otherwise, for the sake
of simplicity, we ignore the damping and assume that the turbulence
cascade is unaffected. As the transfer of energy between
Alfv\'en, slow and fast modes of MHD turbulence is suppressed at the
scales less than the injection scale, we consider different components of MHD cascade independently.

We get the cutoff scale $k_{c}$ by equating the damping rate and
cascading rate $\tau_{k}\Gamma_{d}\simeq1$. Then we check whether
it is self-consistent by comparing the $k_{c}$ with the relevant
scales, e.g., injection scale, mean free path and the ion gyro-scale.

Damping is, in general, anisotropic, i.e., the linear damping (see \cite{YL04} Appendix~A) depends on the
angle $\theta$ between the wave vector ${\bf k}$ and local direction of
magnetic field ${\bf B}$. Unless randomization of $\theta$ is comparable to the cascading rate the damping scale gets angle-dependent\footnote{The fast increase of collisionless damping with $\theta$ was applied to rotating stars, where \cite{Suzuki} showed that the collisionless damping can be a dominant heating source for stellar wind.}. The angle $\theta$ varies because of both the randomization of wave vector $\bf k$ and the wandering of magnetic field lines ( see\cite{YL04} for detail).
\begin{figure}
\includegraphics[ width=0.8\columnwidth]{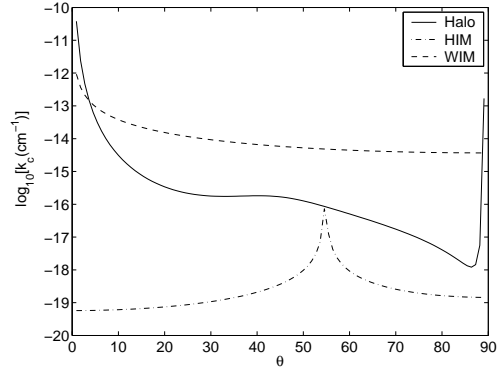}
\caption{Damping scale vs. the angle $\theta$ between ${\bf k}$ and $\bf B$ in halo, HIM and WIM. The peak on the curve of HIM (dashdot line) is smeared out by randomization of both ${\bf k}$ and $\bf B$ \cite{YL04}.}
\end{figure}
 With this input at hand, one can determine the turbulence
damping scales given a medium.
\section{Interactions between turbulence and particles}

Basically there are
two types of resonant interactions: gyroresonance acceleration
and transit acceleration (henceforth TTD). The resonant condition is $\omega-k_{\parallel}v\mu=n\Omega$ ($n=0, \pm1,2...$),
where $\omega$ is the wave frequency, $\Omega=\Omega_{0}/\gamma$
is the gyrofrequency of relativistic particle, $\mu=\cos\xi$,
where $\xi$ is the pitch angle of particles. TTD formally corresponds to $n=0$ and it requires compressible perturbations.  

We employ quasi-linear theory (QLT) to obtain our estimates. If mean magnetic field is larger than the fluctuations at the injection scale, we may say that the QLT treatment we employ defines parallel diffusion.
Taking into account only the dominant interaction at $n=\pm1$, we obtain the Fokker-Planck coefficients (see also \cite{SchAcha, YL04}),

\begin{eqnarray}
\left(\begin{array}{c}
D_{\mu\mu}\\
D_{pp}\end{array}\right)  =  {\frac{\pi\Omega^{2}(1-\mu^{2})}{2}}\int_{\bf k_{min}}^{\bf k_{max}}dk^3\delta(k_{\parallel}v_{\parallel}-\omega \pm \Omega)\nonumber\\
\left[\begin{array}{c}
\left(1+\frac{\mu V_{ph}}{v\zeta}\right)^{2}\\
m^{2}V_{A}^{2}\end{array}\right]\left\{ (J_{2}^{2}({\frac{k_{\perp}v_{\perp}}{\Omega}})+J_{0}^{2}({\frac{k_{\perp}v_{\perp}}{\Omega}}))\right.\nonumber\\
\left[\begin{array}{c}
M_{{\mathcal{RR}}}({\mathbf{k}})+M_{{\mathcal{LL}}}({\mathbf{k}})\\
K_{{\mathcal{RR}}}({\mathbf{k}})+K_{{\mathcal{LL}}}({\mathbf{k}})\end{array}\right]-2J_{2}({\frac{k_{\perp}v_{\perp}}{\Omega}})J_{0}({\frac{k_{\perp}v_{\perp}}{\Omega}})\nonumber\\
\left.\left[e^{i2\phi}\left[\begin{array}{c}
M_{{\mathcal{RL}}}({\mathbf{k}})\\
K_{{\mathcal{RL}}}({\mathbf{k}})\end{array}\right]+e^{-i2\phi}\left[\begin{array}{c}
M_{{\mathcal{LR}}}({\mathbf{k}})\\
K_{{\mathcal{LR}}}({\mathbf{k}})\end{array}\right]\right]\right\} ,
\label{gyro}
\end{eqnarray}
where $\zeta=1$ for Alfv\'{e}n modes and $\zeta=k_{\parallel}/k$
for fast modes, $k_{min}=L^{-1}$, $k_{max}=\Omega_{0}/v_{th}$
corresponds to the dissipation scale, $m=\gamma m_{H}$ is the relativistic
mass of the proton, $v_{\perp}$ is the particle's velocity component
perpendicular to $\mathbf{B}_{0}$, $\phi=\arctan(k_{y}/k_{x}),$
${\mathcal{L}},{\mathcal{R}}=(x\pm iy)/\sqrt{2}$ represent left and
right hand polarization.

The delta function $\delta(k_{\parallel}v_{\parallel}-\omega+n\Omega)$
approximation to real interaction is true when magnetic perturbations
can be considered static%
\footnote{Cosmic rays have such high velocities that the slow variation of the
magnetic field with time can be neglected. %
} . For cosmic rays, $k_{\parallel}v_{\parallel}\gg\omega$
so that the resonant condition is essentially $k_{\parallel}v\mu-n\Omega=0$.
From this resonance condition, we know that the most important interaction
occurs at $k_{\parallel}=k_{\parallel,res}=\Omega/v_{\parallel}$.
This is generally true except for small $\mu$ (or scattering near
$90^{o}$). 

\section{Scattering of cosmic rays}

\subsection{Scattering by Alfv\'{e}nic turbulence}

As we discussed above, Alfv\'{e}n modes are anisotropic, eddies
are elongated along the magnetic field, i.e., $k_{\perp}>k_{\parallel}$.
The scattering of CRs by Alfv\'{e}n modes is suppressed first because
most turbulent energy goes to $k_{\perp}$ due to the anisotropy of
the Alfv\'{e}nic turbulence so that there is much less energy left
in the resonance point $k_{\parallel,res}=\Omega/v_{\parallel}\sim r_{L}^{-1}$.
Furthermore, $k_{\perp}\gg k_{\parallel}$ means $k_{\perp}\gg r_{L}^{-1}$
so that cosmic ray particles have to be interacting with lots of eddies
in one gyro period. This random walk substantially decreases the scattering
efficiency. The scattering by realistic Alfv\'enic turbulence was studied in \cite{YL02}. In case that the pitch angle $\xi$ not close to 0, the analytical result is  

\begin{eqnarray}
\left[\begin{array}{c}
D_{\mu\mu}\\
D_{pp}\end{array}\right]&=&\frac{v^{2.5}\mu^{5.5}}{\Omega^{1.5}L^{2.5}(1-\mu^2)^{0.5}}\Gamma[6.5,k_{max}^{-\frac{2}{3}}k_{\parallel,res}L^{\frac{1}{3}}]\nonumber\\
& &\left[\begin{array}{c}
1\\
m^{2}V_{A}^{2}\end{array}\right],\label{ana}
\end{eqnarray}
where $\Gamma[a,z]$ is the incomplete gamma function. The presence
of this gamma function in our solution makes our results orders of
magnitude larger than those%
\footnote{We compared our result with the resonant term as the nonresonant term
is spurious as noted by \cite{Chan}. %
} in \cite{Chan} for the most of energies considered. However,
the scattering frequency,

\be
\nu=2D_{\mu\mu}/(1-\mu^{2}),\label{nu}
\ee
are still much smaller
than the estimates for isotropic and slab model \cite{YL02}. As the anisotropy of the Alfv\'{e}n modes is increasing with the
decrease of scales, the interaction with Alfv\'{e}n modes becomes
more efficient for higher energy cosmic rays. When the Larmor radius
of the particle becomes comparable to the injection scale, which is
likely to be true in the shock region as well as for very high energy cosmic
rays in diffuse ISM, Alfv\'{e}n modes get important.


It's worthwhile to mention the imbalanced cascade of Alfv\'{e}n modes
\cite{CLV02}. Our basic assumption above was that Alfv\'{e}n modes were
balanced, meaning that the energy of modes propagating one way was
equal to that in opposite direction. In reality, many turbulence sources
are localized so that the modes leaving the sources are more energetic
than those coming toward the sources. The energy transfer in the imbalanced
cascade occurs at a slower rate, and the Alfv\'{e}n modes behave
more like waves. The scattering by these imbalanced Alfv\'{e}n modes could be more efficient. However, as the actual 
degree of anisotropy of imbalanced
cascade is currently uncertain, and the process will be discussed elsewhere\footnote{Preliminary results by one of us show that the inertial range
over which the degree of anisotropy of imbalanced turbulence is small
is very limited.}. 

\subsection{Scattering by fast modes}

The contribution from slow modes is no more than that by Alfv\'{e}n modes since the slow modes have the similar anisotropies and scalings.
More promising are fast modes, which are isotropic. With fast
modes there can be both gyroresonance
and transit-time damping (TTD).

TTD happens due to the resonant
interaction with parallel magnetic mirror force.
The advantage
of TTD is that it doesn't have a distinct resonant scale associated
with it. TTD is thus an alternative to scattering of low energy CRs which
Larmor radii are below the damping scale of the fast modes. Moreover,
we shall show later that TTD can contribute substantially to cosmic ray acceleration (also
known as the second order Fermi acceleration). This can be crucial in
some circumstances, e.g., for $\gamma$ ray burst (\cite{gammaray}), and acceleration of charged particles \cite{YL03}. Different from gyroresonance, the resonance function of TTD is broadened even for CRs with small pitch angles. The formal resonance peak $k_{\parallel}/k=V_{ph}/v_{\parallel}$ favors quasi-perpendicular modes. However, these quasi-perpendicular modes cannot form an effective mirror to confine CRs because the gradient of magnetic perturbations along the mean field direction $\nabla_{\parallel}\mathbf{B}$ is small. Thus the resonance peak is weighted out and the Breit-Wigner-type \cite{SchAcha} resonance function should be adopted.  

Here we apply our analysis to the various phases of ISM. In low $\beta$ medium, both collisionless and collisional damping increases with $\theta$ unless $\theta$ is close to
$\pi/2$. This
means that those quasi-parallel fast modes are least damped. For these modes the argument for the Bessel function
in Eq.(\ref{gyro}) is $k_{\perp}\tan\xi/k_{\parallel,res}<1$
unless $\xi$ is close to $90^{o}$. So we can take advantage of
the anisotropy of the damped fast modes and use the asymptotic form
of Bessel function for small argument $J_{n}(x)\simeq(x/2)^{n}/n!$
to obtain the corresponding analytical result for this case:
\begin{eqnarray}
\left[\begin{array}{c}
D_{\mu\mu}\\
D_{pp}\end{array}\right]=\frac{\pi(\Omega v\mu)^{0.5}(1-\mu^{2})}{2L^{0.5}}\nonumber \\
\left[\begin{array}{c}
(1-(\left(\frac{k_{\perp,c}}{k_{\parallel,res}}\right)^{2}+1)^{-\frac{7}{4}})/7\\
(1-(\left(\frac{k_{\perp,c}}{k_{\parallel,res}}\right)^{2}+1)^{-\frac{3}{4}})m^{2}V_{A}^{2}/3\end{array}\right]
\label{fastlb}\end{eqnarray}

\begin{figure}
\leavevmode
\includegraphics[ width=0.48\columnwidth]{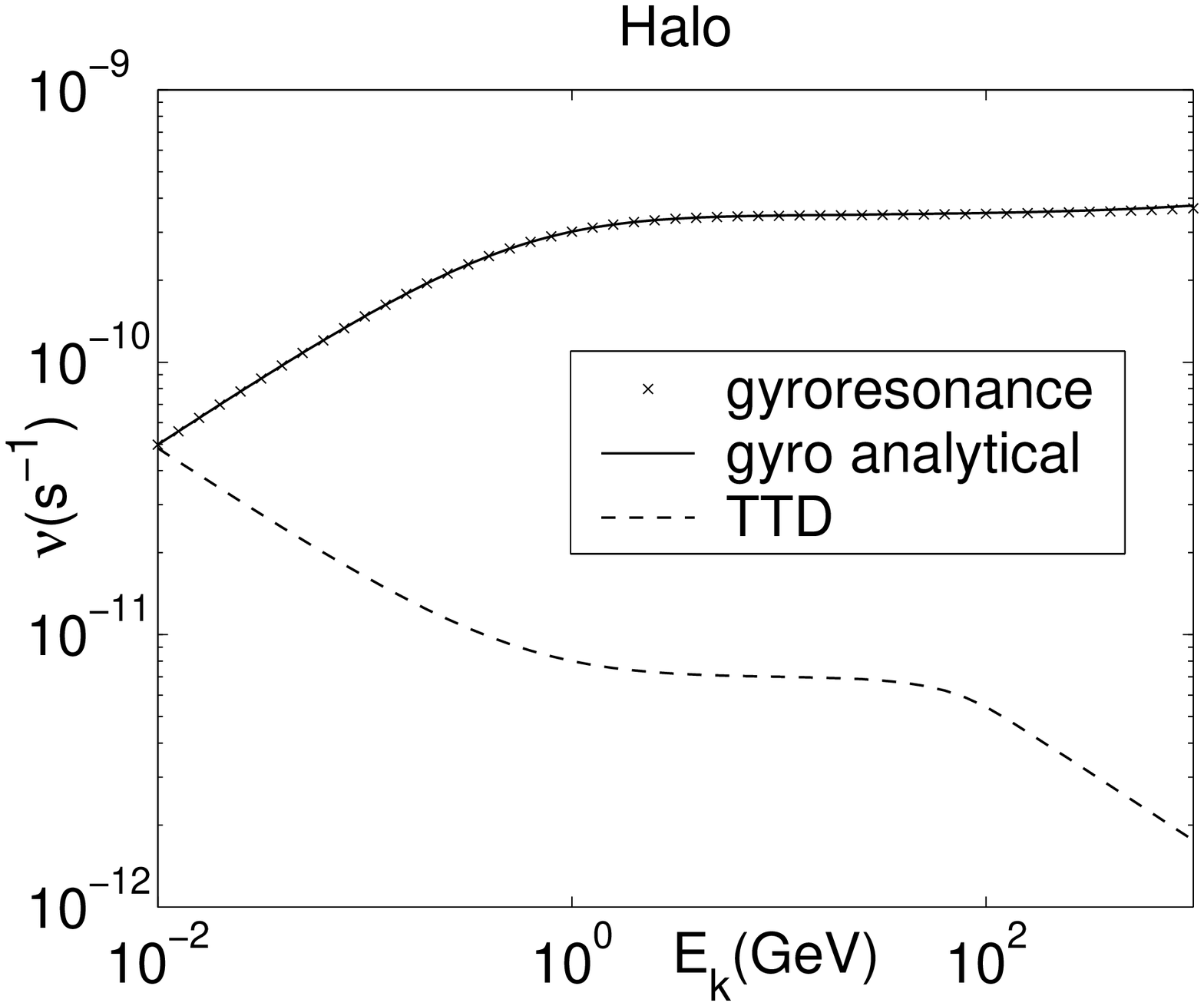} \hfil
\includegraphics[  width=0.48\columnwidth]{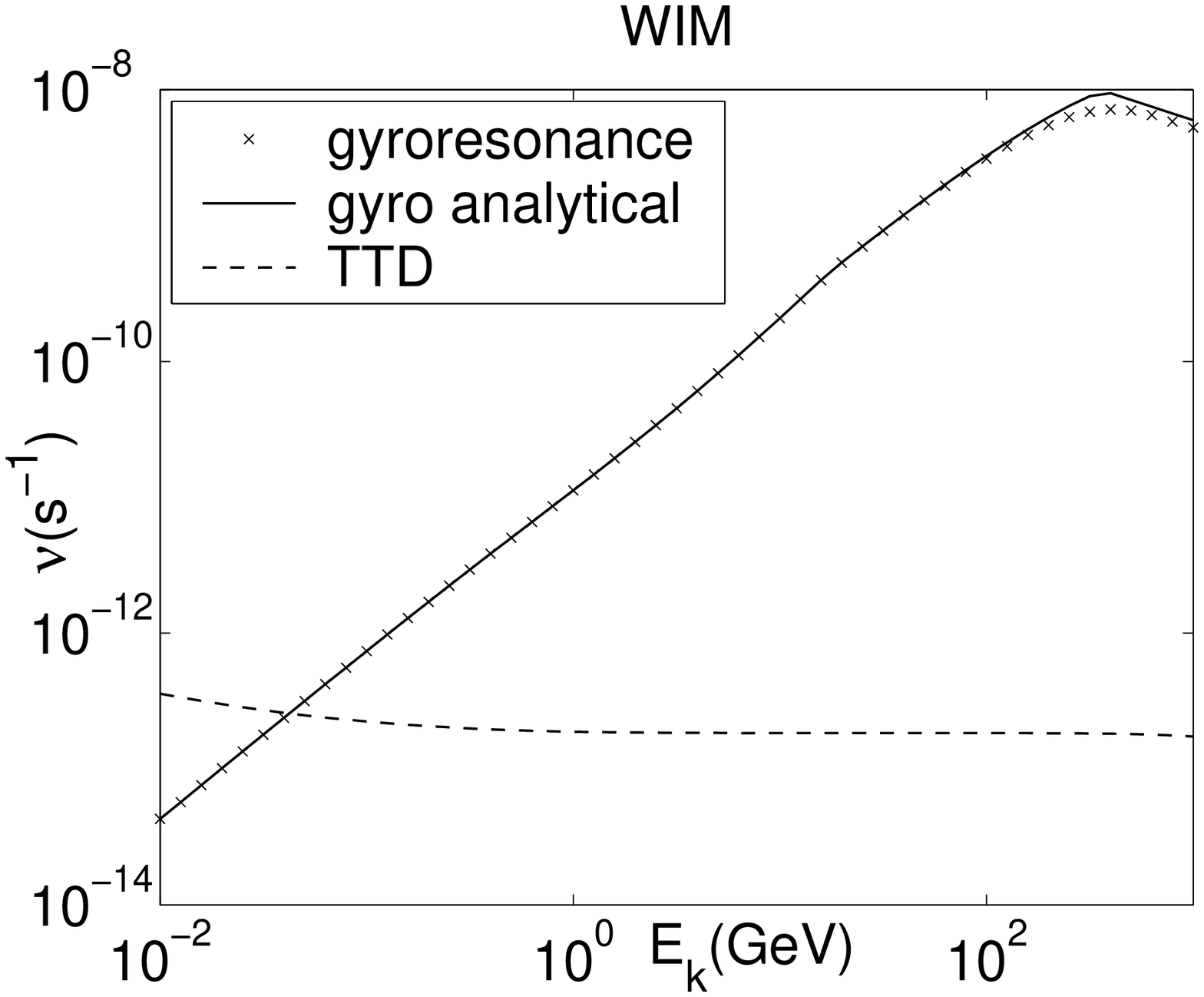} 
\caption{Scattering frequency $\nu$ given by Eq.(\ref{nu}) vs. the kinetic energy $E_{k}$ of cosmic rays (a) in halo,  (b) in WIM. The 'x' lines refer to scattering by gyroresonance and solid lines are the corresponding analytical results given by Eq.(\ref{fastlb}). 
The dashed line are the results for TTD (from \cite{YL04}).}
\end{figure}

In high $\beta$ medium and partially ionized medium, fast modes are subjected to severe damping and truncated at scale larger than the resonant scale of CRs, $\sim \Omega/c$. The only available scattering mechanism is TTD, which can still provide much more efficient scattering for CRs comparing with Alfv\'en modes (see \cite{YL04} for detail).   

A special case is that the cosmic rays propagate nearly perpendicular
to the magnetic field, so called the $90^{o}$ scattering problem.
It should be noted that with resonance broadening associated with the   turbulence, the requirement $v_{\parallel}=V_{ph}/\cos\theta>V_{ph}$ is relieved. The contribution from TTD thus becomes dominant for sufficiently small $\mu$. However, since quasi-linear approximation is not accurate in this regime, proper calculation should be carried out with non-linear effects taken into account \cite{Owens, Gold}.

\begin{figure}
\includegraphics[ width=0.48\columnwidth]{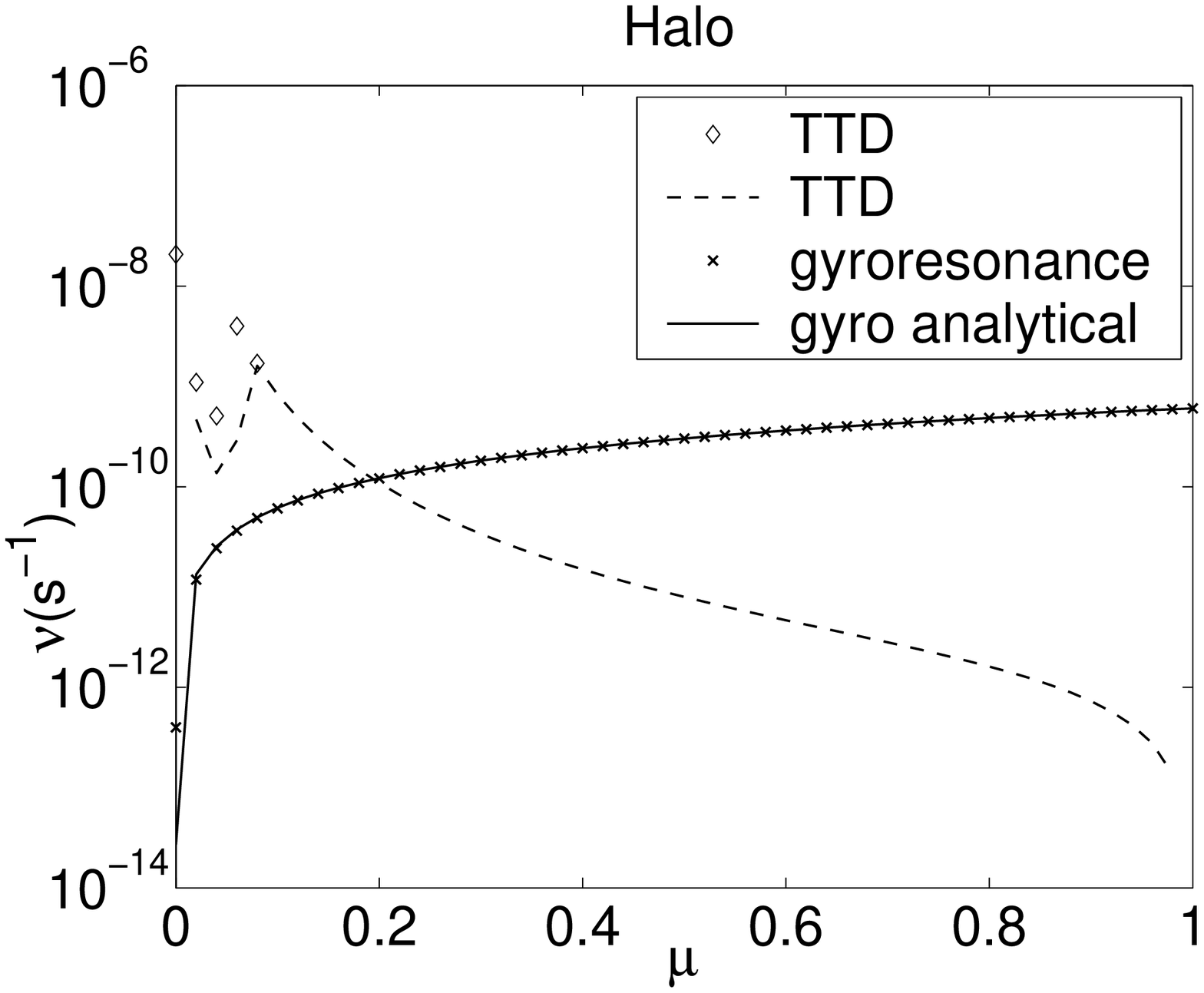} \hfil 
\includegraphics[ width=0.48\columnwidth]{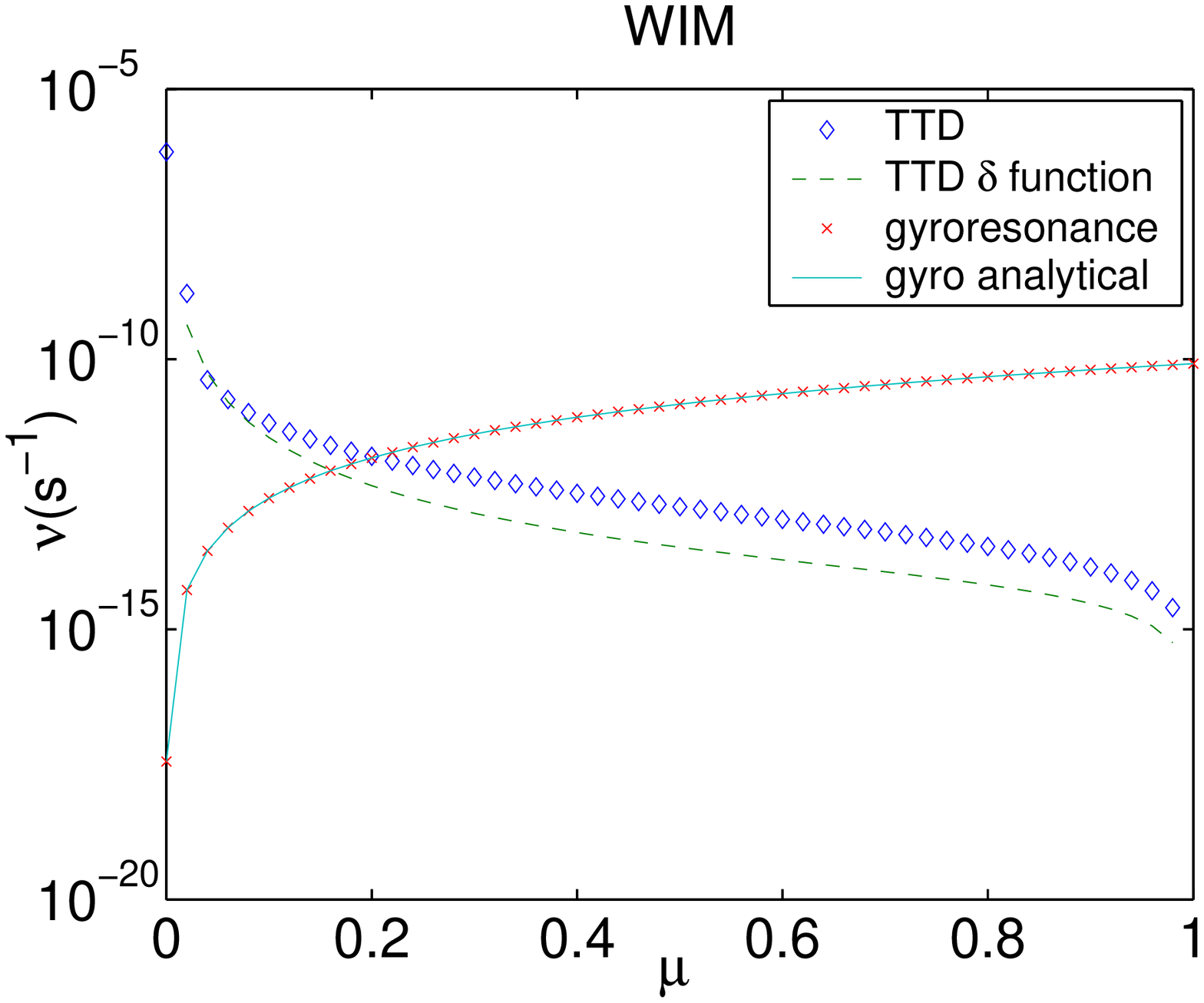} 
\caption{The scattering frequency $\nu$ vs. cosine of pitch angle $\mu$ of 1GeV CR (a) in halo, (b) in WIM.  The 'x' lines refer to scattering by gyroresonance and solid lines are the corresponding analytical results given by Eq.(\ref{fastlb}). 
The dashed and diamond lines are the results for TTD adopting $\delta$ function and Breit-Wigner function respectively (from \cite{YL04}).} 
\end{figure}

\section{Cosmic ray self confinement by streaming instability}

When cosmic rays stream at a velocity much larger than Alfv\'{e}n
velocity, they can excite by gyroresonance MHD modes which in turn scatter
cosmic rays back, thus increasing the amplitude of the resonant mode. This 
runaway process is known as streaming instability. It was claimed
that the instability could provide confinement for cosmic rays with
energy less than $\sim 10^2$GeV \cite{Cesar}. However, this was calculated
in an ideal regime, namely, there was no background MHD turbulence.
In other words, it was thought that the
self-excited modes would not be appreciably damped in fully ionized gas. 

This is not true for turbulent medium, however. \cite{YL02}
pointed out that the streaming instability is partially suppressed
in the presence of background turbulence\footnote{The fast cascade
induces fast non-linear damping of MHD turbulence. Essentially the damping
of Alfv\'enic turbulence happens in one eddy turnover time for large eddies.
This effect was invoked in \cite{Pohl} to explain the
small transversal size of X-ray filaments observed (\cite{Bamba})}. More recently, detailed calculations of the 
streaming instability in the presence of background Alfv\'enic turbulence 
were presented in \cite{Farmer}.

For interaction with fast modes, it happens at the rate 
$\tau_k\sim (k/L)^{-1/2}V_{ph}/V^2$ (see Eq.(\ref{fdecay})). 
By equating the growth rate \cite{Longair},
\begin{equation}
\Gamma(k)=\Omega_0\frac{N(\geq E)}{n_{p}}(-1+\frac{v_{stream}}{V_{A}})
\label{instability}
\end{equation}
and the damping rate 
Eq.(\ref{fdecay}), we can find that the streaming instability
is only applicable for particles with energy less than
\begin{equation}
\gamma_{max}\simeq1.5\times 10^{-9}[n_{p}^{-1}(V_{ph}/V)(Lv\Omega_0/V^2)^{0.5}]^{1/1.1},
\end{equation}
which for HIM, provides $\sim 20$GeV if taking the injection speed to be $V\simeq 25$km/s. 

One of the most vital cases for streaming instability is that of cosmic
ray acceleration in strong MHD shocks. Such shocks produced by supernovae 
explosions scatter cosmic rays by the postshock turbulence and by preshock
magnetic perturbations created by cosmic ray streaming. The perturbations
of the magnetic field may be substantially larger than the regular magnetic
field. The corresponding nonlinear growth rate is (\cite{Ptuskin}):
\begin{equation}
\Gamma(k)\simeq\frac{ak \epsilon U^3}{c V_A(\gamma_{max}^a-(1+a)^{-1})(kr_{g0})^a(1+A_{tot}^2)^{(1-a)/2}},
\label{shock}
\end{equation}
where $\epsilon$ is the ratio of the pressure of CRs at the shock and the upstream momentum flux entering the shock front, $U$ is the shock front speed, $a-4$ is the spectrum index of CRs at the shock front, $r_{g0}=c/\Omega_0$, $A_{tot}$ is the dimensionless amplitude of the random field $A_{tot}^2=\delta B^2/B_0^2$. 

Magnetic field itself is likely to be amplified through an inverse 
cascade of magnetic energy at which perturbations created at a particular
$k$ diffuse in $k$ space to smaller $k$ thus inducing inverse cascade. 
As the result the magnetic perturbations at smaller $k$ get larger than the 
regular field. As the result, even if the instability is suppressed
for the growth rate given by eq. (\ref{instability}) it gets efficient
due to the increase of perturbations of magnetic field stemming from the
inverse cascade.

Whether or not the streaming instability is efficient in scattering
accelerated cosmic rays back depends on whether the growth rate of
the streaming instability is larger or smaller than the damping rate.
The precise picture of the process depends on yet not completely
clear details of the inverse cascade of magnetic field. 
If, however, we assume that the small scale driving provides at 
the scales of interest isotropic turbulence the nonlinear damping
happens on the scale of one eddy turnover time. Assuming the shock front speed $U$ is low, we attain the maximum energy of particles accelerated in the shock by equating the growth rate of the instability Eq.(\ref{shock}) to the damping rate due to turbulence cascade Eq.(\ref{fdecay}): 
\begin{equation}
\gamma_{max}\simeq (\frac{a\epsilon (LeB_0)^{0.5}U^3}{m^{0.5}V^{2}c^2})^{1/(0.5+a)}.
\end{equation}
From this we can estimate $\gamma_{max}\simeq 2\times 10^7 (t/kyr)^{-9/4}$ in HIM.

As shown in the Eq.(\ref{instability}) and (\ref{shock}), the
growth rate depends on the CRs' density. In those regions where high
energy particles are produced, e.g., shock fronts in ISM, $\gamma$ ray burst, SN, the
streaming instability is more important. 

\section{Summary}

In the paper above we characterized interaction of cosmic rays with 
balanced interstellar turbulence driven at a large scale. Our results 
can be summarized as follows:

1. fast modes provide the dominant contribution to scattering of cosmic
rays in different phases of interstellar medium
provided that the turbulent energy is injected at large scales.
This happens in spite of the fact that the fast modes are more
subjected to damping compared to Alfv\'en modes. 

2. As damping of fast modes depends on the angle between
the magnetic field and the wave direction of propagation,
we find that field wondering determined by Alfv\'en modes affects
the damping of fast modes. At small scales the 
anisotropy of fast mode damping
makes gyroresonant scattering within slab approximation 
applicable. At larger 
scales where damping is negligible the isotropic gyroresonant 
scattering approximation is applicable.

3. Transient time damping (TTD) provides an important means of cosmic ray transport. Use of $\delta$ function resonance entails errors, and therefore, resonance broadening is essential for TTD. Our study shows that it is vital for low energy and large pitch angle scattering. And it dominates scattering of cosmic rays in HIM and the partially 
ionized interstellar gas where fast modes are severely damped.

4. Streaming instability is subjected to non-linear damping
 due to the interaction of 
the emerging magnetic perturbations with the surrounding turbulence.
The energy at which the streaming instability is suppressed
depends on whether on the inverse cascade of magnetic energy
as the instability gets more easily excited for low energy particles.

\end{document}